\documentstyle[preprint,aps,prb]{revtex}
\begin{document}
\preprint{Phys.\ Rev.\ B {\bf 51}, 6 999 (1995)}
\draft
\title{The Nature of Thermopower in Bipolar Semiconductors}

\author{Yu.\ G. Gurevich\cite{byline} and O. Yu.\ Titov}
\address{Centro de Investigaci\'{o}n
y de Estudios Avanzados del Instituto Polit\'{e}cnico Nacional, Apartado
Postal 14--740,\\ M\'{e}xico 07000, Distrito Federal, M\'{e}xico}
\author{G. N. Logvinov}
\address{Teacher Training University, Avenue Krivonosa 2,
	Ternopol 282009, Ukraine}
\author{O. I. Lyubimov}
\address{Kharkov State University,
	Square Svobodi 4, Kharkov 310077, Ukraine.}
\date{17 February 1994}
\maketitle
\begin{abstract}
The thermoemf in bipolar semiconductors is calculated. It is shown that it is
necessary to take into account the nonequilibrium distribution of
electron and hole concentrations (Fermi quasilevels of the electrons and
holes). We find that electron and hole electric conductivities of
contacts of semiconductor samples with connecting
wires make a substantial contribution to thermoemf.
\end{abstract}
\section*{Introduction}
In calculating thermo-emf and explaining its nature, it is common to
consider at first the unipolar case.\cite{Anselm78}
The physical transparency of this phenomenon and the clearness of
calculation  lead to the following paradoxical results: in the case of a
bipolar medium the situation seems to be equally obvious, and so the
same calculation scheme is used.\cite{Anselm78}
 The aim of this paper is to show that the situation changes in bipolar
media in principle, from the point of view of both the physics of the
proceeding processes and the methods of thermo-emf calculations.

Let us begin with the unipolar situation. Dating back to the paper of
Thomson\cite{Thompson1882} in 1856 the theory of the origin of the
thermo-emf in this case has found its most precise description in the
publications of Seeger and Kaydanov and Nuramski.\cite{Kaydanov81} It is
necessary to connect the electric circuit (Fig.\ \ref{fig:single}) for
the determining thermo-emf in the case of the absence of electric
current (broken circuit). A semiconductor sample whose thickness is
$2a$ ($ -a \le x \le a$) contacts with a heater with temperature $T_1$ on
the surface $x=-a$ and with a cooler with temperature $T_2$ on the
surface $x=+a$. The connecting wires are made of the same material
(metal) and are hooked up to the terminals of
a measuring compensating circuit which allows us to measure the difference
of voltage $\varphi_b$ and $\varphi_c$ in the absence of an electric current.
Leads of connecting wires have equal temperatures [for example,
$T^*=(T_1 + T_2) / 2$] at points $b$ and~$c$. As follows from Ohm's law,
for a closed electric circuit $V=j(R+r)$ ($V$ is the emf of the power
source, $r$ is source's resistance, $R$ is the r
esistance of the external load, and $j$ is the density of the electric
current). If $R\to\infty$ (broken circuit), then the density of electric
current $j\to0$ and $V=jR=\varphi_c-\varphi_b$.\cite{GY}

Since the electron's chemical potentials $\mu_n(x)$ are equal in the
points $b$ and~$c$, then \begin{equation}
\varphi_c-\varphi_b = (\tilde\mu_n^b-\tilde\mu_n^c)/e,
\label{eq:first}
\end{equation}
where $\tilde \mu_n(x)=\mu_n(x) -e\varphi(x)$ is an electrochemical
potential. At the same time, \begin{equation}
\tilde\mu_n^b-\tilde\mu_n^c=\int_c^b{d\tilde\mu_n(x) \over dx}\,dx.
\label{eq:second}
\end{equation}

As is well known, the expression for the density of the electric current
is of the form\cite{Anselm78} \begin{equation}
 {\bf j}_n=\sigma_n(x)\left(\nabla{\tilde\mu_n(x) \over e}
	     - \alpha_n(x)\nabla T\right),
\label{eq:third}
\end{equation}
where $\sigma_n(x)$ is the electric conductivity and $\alpha_n(x)$ is
the thermoelectric power. In the absence of an electric current
$d\tilde\mu_n(x) / dx = e \, \alpha_n(x)dT/dx$. Let us emphasize that
this correlation is correct all over the circuit, where $\sigma_n(x)\ne
0$, e.g., outside the region [$b$,$c$], where $\sigma_n(x)\to0$ (this
condition provides ${\bf j}_n={\bf 0}$ everywhere in the circuit).
Therefore the expression (\ref{eq:second}) can be rewritten in the form
\[
\tilde\mu_n^b-\tilde\mu_n^c=\int_c^b e\,\alpha_n(x) {dT \over dx}\,dx=
\int_{T^*}^{T_2}e\,\alpha_M\,dT +
\int_{T_2}^{T_1}e\,\alpha_n\,dT
 +
\int_{T_1}^{T^*}e\,\alpha_M\,dT,
\]
where $\alpha_n$ and $\alpha_M$ are the values of $\alpha_n(x)$ in the
semiconductor and in the connecting wires. Finally we have for the
difference $\varphi_c-\varphi_b$, which coincides with the thermo-emf
$V$ of the broken circuit, \begin{equation}
V=\varphi_c - \varphi_b = (\alpha_n - \alpha_M)(T_1 - T_2).
\label{eq:unipol_V}
\end{equation}

To simplify the calculations we have assumed that $T_1-T_2 \ll T^*$
which does not restrict the generality. Since usually $| \alpha_M | \ll
| \alpha_n |$, then \begin{equation}
V = \alpha_n (T_1 - T_2).
\label{eq:simp_unipol_V}
\end{equation}
The last expression coincides formally with
$\int_{-a}^{+a}\nabla[\varphi-(\mu_n / e)]\,dx$.

Thus the general scheme of calculating the thermo-emf comes to the
following for a unipolar broken circuit. The gradient of the
electrochemical potential $\nabla\tilde\mu_n(x)$ is
obtained by setting expression (\ref{eq:third}) equal to zero. Then
integration of $\nabla\tilde\mu_n(x)$ in the anticlockwise direction
(together with multiplying by $e^{-1}$) gives the value of the
thermo-emf. It is important to emphasize that, as above,  the extreme
points of the circuit ($b$ and $c$) would have the same temperatures.

It follows from (\ref{eq:second}), in a unipolar medium, that even if
the condition of quasineutrality is not fulfilled and nonequilibrium
electrons (this notion will be defined more exactly below) arise, the
gradient of concentration of these carriers $n$ does not give a
contribution to the thermo-emf [the summand $(\partial\mu_n / \partial
n)\nabla n$ disappears from the expression for~$V$]. This argument
confirms the following idea of Ioffe:\cite{Ioffe60} the potential
difference created by the bulk concentration's gradient is compensated
by the contact potential difference (``diffusion voltages'') on the
boundaries $+a$ and $-a$.

Let us note in conclusion that the obviousness of the above scheme is
lost when thermo-electric current flows in a closed circuit [$j_n \ne
0$, see Eq.~(\ref{eq:temp_over_conditions})].

\section{Bipolar semiconductors. Traditional approach}
\label{sec:bsta}
The system of equations for electrons and holes are analogous to
(\ref{eq:third}):\cite{Anselm78,BBG}
\begin{eqnarray}
	{\bf j}_n &=& \sigma_n(x) \left[+ \nabla{\tilde\mu_n(x) \over e} -
		  \alpha_n(x) \nabla T \right],\nonumber \\
	{\bf j}_p &=& \sigma_p(x) \left[- \nabla{\tilde\mu_p(x) \over e} -
		  \alpha_p(x) \nabla T \right].
\label{eq:bpt_currents}
\end{eqnarray}

Here ${\bf j}_p$ is the density of the hole electric current, $\sigma_p(x)$ and
$\alpha_p(x)$ are the electric conductivity and thermoelectric power of
the holes [$\alpha_n(x)$ and $\alpha_p(x)$ have opposite signs],
\begin{equation}
	\tilde\mu_p = - \varepsilon_g - \tilde\mu_n = \mu_p + e\varphi
\label{eq:bpt_el-ch_pot}
\end{equation}
is the hole electrochemical potential,\cite{Anselm78} and
$\varepsilon_g$ is the band gap. It is important to emphasize that in
expressions~(\ref{eq:bpt_currents})
$\nabla\mu=(\partial\mu / \partial T)\nabla T$ usually. The full current
${\bf j}$ is equal to \begin{equation}
	{\bf j} ={\bf j}_n + {\bf j}_p = 0,
\label{eq:bpt_full_current}
\end{equation}
if a bipolar semiconductor is represented as in Fig.\ \ref{fig:single}.

It is easy to obtain $\nabla[\tilde\mu_n(x) / e]$ from
(\ref{eq:bpt_full_current}) taking into account (\ref{eq:bpt_currents})
and knowing that $\nabla\tilde\mu_p=-\nabla\tilde\mu_n$ [see
(\ref{eq:bpt_el-ch_pot})],
\[
	\nabla \left[ {\tilde\mu_n(x) \over e} \right] =
		{1 \over \sigma(x)}
		\left[ \sigma_n(x) \alpha_n(x) + \sigma_p(x) \alpha_p(x)
		\right]
		\nabla T,\qquad
		\sigma(x) = \sigma_n(x) + \sigma_p(x).
\]
Then we find the thermo-emf $V$ (it is assumed that
$\alpha_{n,p} \gg \alpha_M$ and $T_1-T_2  \ll  T^*$ as above):
\begin{equation}
	V = \int_c^b \nabla \left[{\tilde\mu_n(x) \over e}\right] \, dx =
	    {1 \over \sigma}
	    \left( \sigma_n\alpha_n + \sigma_p \alpha_p \right)
	    (T_1 -T_2).
\label{eq:bpt_V}
\end{equation}

This expression, especially the method of the calculation, causes
serious objections. Really, the electron concentration and hole
concentration should be lower in the heating lead in the stationary
state due to thermodiffusion if bulk and surface recombin
ations are absent. In contrast, these concentrations should increase in
the cooling lead. On the one hand, this
causes the appearance of appreciable diffusion currents\cite{GM} in
expression
(\ref{eq:bpt_currents}) [$\nabla\mu_n=(\partial \mu_n / \partial
n)\nabla n$].
On the other hand, it leads to a violation of relation
(\ref{eq:bpt_el-ch_pot}). Two Fermi
quasilevels $\tilde\mu_n$ and $\tilde\mu_p$ arise instead of a single level
of the  electrochemical potential, and as a result
$\left|\nabla\tilde\mu_n\right| \ne \left|\nabla\tilde\mu_p\right|$. In
this case the procedure described in the beginning of this
section becomes incorrect, because the single common ``gradient of
electrochemical potential'' of electrons and holes is absent. Moreover,
if bulk and surface recombinations are absent then both partial currents
${\bf j}_n$ and ${\bf j}_p$ should be equal
 to zero, not only the full
current ${\bf j}$ (${\bf j}_n+{\bf j}_p={\bf 0}$). As a result we have
two equations (${\bf j}_n={\bf 0}$, ${\bf j}_p={\bf 0}$) for both
thermoelectric fields $\nabla(\tilde\mu_n / e)$ and $\nabla(\tilde\mu_p
/ e)$ instead of one equation (\ref{eq:bpt_ful l_current}).
One more problem arises when bulk and surface recombinations take place:
the correct determination of electron and hole equilibrium
concentrations.\cite{GM}

Finally, the question remains how to obtain the thermo-emf in this case,
and which physical phenomena determine its value. The answer to the
first of these questions has been given,\cite{GY} where the general
scheme was proposed for calculation of an emf of any nature.
It follows from this paper that\cite{GM}
\begin{equation}
	V = \oint\left(
		  {\sigma_n \over e\sigma} {d\tilde\mu_n \over dx} -
		  {\sigma_p \over e\sigma} {d\tilde\mu_p \over dx}
		  \right) dl -
	    \oint\left(
		  {\sigma_n\alpha_n \over \sigma} +
		  {\sigma_p\alpha_p \over \sigma}
		  \right) {dT \over dx} dl,
\label{eq:temf_of_any_nature}
\end{equation}
where integration is carried out clockwise. Let us note that expression
(\ref{eq:temf_of_any_nature}) is always correct (for a broken circuit
just as in the case of a flowing thermoelectric current). The second
item in (\ref{eq:temf_of_any_nature}) coinci
des with the expression (\ref{eq:bpt_V}). The first item in
(\ref{eq:temf_of_any_nature}) vanishes (it is equal to zero identically
in the unipolar case) and expression (\ref{eq:temf_of_any_nature}) turns
into (\ref{eq:bpt_V}) if electrons and holes have a single,
common electrochemical potential level. Correlation
(\ref{eq:bpt_el-ch_pot}) does not hold if electrons and holes flow from
the hot lead to the cool one and electron and hole
Fermi quasilevels appear. In this situation the first item in
(\ref{eq:temf_of_any_nature}) differs from zero and the gradients of
concentrations and corresponding diffusion currents contribute to the
thermo-emf.

\section{Thermoelectric phenomena in bipolar mediums}
Let us go on to the description of an approach to exploration of the
thermo-emf, which does not involve either the contradictions or the
incorrectness pointed out above. Note that some aspects of this approach
have previously been expounded.\cite{GM,GL}

Let us restrict ourselves to the first case
\begin{equation}
	T(x) = T^* - {\Delta T \over 2a}x,
\label{eq:temp_profile}
\end{equation}
for simplicity. Here $T_1$ is the heater temperature at the point
$x=-a$, $T_2$ is the condenser temperature at the point $x=+a$,
$T^*=(T_1+T_2) / 2$, and $\Delta T = T_1 - T_2$.

The condition when the temperature field of the quasiparticles
(electrons, holes, and phonons) is common and is a linear function of
coordinates has been obtained earlier.\cite{BBG} Let us assume that
\begin{equation}
	\Delta T  \ll  T^*.
\label{eq:temp_condition}
\end{equation}

Let an arbitrary semiconductor be defined by the function $\mu_n^0(T^*)$
which is founded from the condition of electroneutrality.\cite{Anselm78}
Then $\mu_n^0$ becomes a function of the coordinate $\mu_n^0 \biglb(
T(x) \bigrb)$ in the temperature field ( \ref{eq:temp_profile}), and
\begin{equation}
	\mu_n^0(x) = \mu_n^0(T^*) + \delta\mu_n^0(x),\quad
	\delta\mu_n^0(x) = - \frac{\partial \mu_n^0(T^*)}{\partial T^*}
		\Delta T{x \over 2a},\qquad
	\left|\delta\mu_n^0\right|  \ll  \left|\mu_n^0(T^*)\right|.
\label{eq:temp_over_conditions}
\end{equation}

Function (\ref{eq:temp_over_conditions}) gives, uniquely,
the concentration distribution in the sample:
\begin{eqnarray}
n_0(x) &=& n_0(T^*) + \delta n_0(x),\nonumber \\
n_0(T^*) &=& \gamma_n(T^*)\exp\left[{\mu_n^0(T^*) \over T^*}\right],\nonumber
\\
\delta n_0(x) &=& n_0(T^*) \left[ \left( {\mu_n^0(T^*) \over T^*}
	     - {{3 \over 2}} \right) {{\Delta T \over T^*}}{{x \over 2a}} +
	     {\delta\mu_n^0 \over T^*}\right].
\label{eq:temp_consentrations}
\end{eqnarray}

Here $\gamma_n(T)\propto T^{3/2}$ is the density of states at the bottom
of the conduction band.

If we introduce (here it was assumed that the energy gap $\varepsilon_g$
is independent of temperature)
\[
\mu_p^0(T^*)=-\varepsilon_g - \mu_n^0(T^*),
\]
then we  can write
\begin{equation}
\mu_p^0(x)=\mu_p^0(T^*) + \delta\mu_p^0(x),\qquad
\delta\mu_p^0=-\delta\mu_n^0,
\label{eq:temp_chem_pot}
\end{equation}
analogously to (\ref{eq:temp_over_conditions}). Then the hole concentration is
\begin{eqnarray}
p_0(x) &=& p_0(T^*) + \delta p_0(x),\nonumber \\
p_0(T^*) &=& \gamma_p(T^*)
	\exp\left[{\mu_p^0(T^*) \over T^*}\right],\nonumber \\
\delta p_0(x) &=& p_0(T^*)\left[\left({{\mu_p^0(T^*) \over T^*}} -
	{{3 \over 2}}\right) {\Delta T \over T^*} {x \over 2a} +
	{\delta\mu_p^0 \over T^*}\right],
\label{eq:temp_con_of_holes}
\end{eqnarray}
where $\gamma_p(T)\propto T^{3/2}$ is the density of states at the top
of the valence  band.

Let us note that condition (\ref{eq:temp_condition}) is not sufficient
for the correctness of formulas (\ref{eq:temp_consentrations}) and
(\ref{eq:temp_con_of_holes}) in contrast
to (\ref{eq:temp_over_conditions}) and (\ref{eq:temp_chem_pot}). The
additional conditions \[
|\delta\mu_{n,p}^0|  \ll  T^*,\qquad
{\Delta T \over T^*}  \ll  {T^* \over \max[|\mu_n^0|, |\mu_p^0|]}
\]
are necessary.

It is important to emphasize that $n_0(x)$ and $p_0(x)$ are not
``equilibrium'' concentrations (see the Appendix). Inverted commas are
used here because it is impossible to use the term ``equilibrium,''
strictly speaking, in the presence of a temperature field
(\ref{eq:temp_profile}).

The situation becomes nonequilibrated in the authentic sense (see the
Appendix) when the gradient of electrochemical potential becomes nonzero
because of taking into account the terms $\alpha_n\nabla T$ and $\alpha_p\nabla
T$. Let us examine this situation, assuming that bulk and surface
recombinations are absent, and under the condition of a broken circuit (as in
Sec.~\ref{sec:bsta}).

In this case the stationary distributions of concentrations
\[
n(x)=n_1(x) + \delta n,\qquad p(x)=p_1(x) + \delta p
\]
and electric potential
\[
\varphi(x)=\delta \varphi_1(x) + \delta\varphi
\]
are described by the system of equations ${\bf j}_n={\bf 0}$, ${\bf
j}_p={\bf 0}$ and \[
{d^{\,2}(\delta\varphi) \over dx^2} = 4 \pi e (\delta n - \delta p)
\]
[expressions for $n_1(x)$, $p_1(x)$, and $\delta\varphi_1(x)$ are contained in
(\ref{eq:3d}) and (\ref{eq:quasi_dfi1_dn1})].

Let us note [see (\ref{eq:bpt_currents})] that the quantities [$\mu_n^{(1)}$
($\mu_p^{(1)}$) are the ``equilibrium'' electron (hole) chemical
potentials, see (\ref{eq:3d}) and (\ref{eq:quasi_dfi1_dn1})]
\[
\mu_n = \mu_n^{(1)} + \delta \mu_n,\qquad
\mu_p = \mu_p^{(1)} + \delta \mu_p,
\]
\[
\delta n = {n_0(T^*) \over T^*}\delta\mu_n,\qquad
\delta p = {p_0(T^*) \over T^*}\delta\mu_p,
\]
which are contained in the expressions for ${\bf j}_n$ and ${\bf j}_p$ are
not already connected by expression (\ref{eq:temp_chem_pot}), e.g., they
are compatible with Fermi quasilevels. As for quantities $\alpha_n$ and
$\alpha_p$, depending on $\mu_n$ and $\mu_p$
respectively,\cite{Anselm78} they are determined by the
quantities $\mu_n^0(T^*)$ and $\mu_p^0(T^*)$, since we carry out
all calculations up to the members of order $(\Delta T / T^*)$. Thus
$\delta n_0$, $\delta p_0$, $\delta n_1$, $\delta p_1$, and
$\delta\varphi_1$
($\delta\mu_n^0$, $\delta\mu_p^0$, $\delta\mu_n^{(1)}$, $\delta\mu_p^{(1)}$,
and $\delta\varphi_1$) are not incorporated into the system of equations
for finding $\delta n$, $\delta p$, and $\delta\varphi$ (or
$\delta\mu_n$, $\delta\mu_p$, and $\delta\varphi$).
As a result, the determining of unknown quantities needs no calculations
presented by formulas
(\ref{eq:te mp_over_conditions})--(\ref{eq:temp_con_of_holes}) and
(\ref{eq:chem_pot_cond})--(\ref{eq:quasi_dfi1_dn1}). We shall recall,
however, that these
calculations are necessary if we have to take into account recombination.
Conditions (\ref{eq:surface_con_for_curr}) and $\left.
\varphi\right|_{x=0}=0$
are used for determining integration constants.

Then
\[
\delta\varphi = A_1 (e^{\lambda x} -1) + A_2(e^{-\lambda x} - 1) +
		\alpha\Delta T { x \over 2a},
\]
\[
\delta\mu_n = + e \, \delta\varphi + e \, \alpha_n T(x) + e \, C_1,
\]
\begin{equation}
\delta\mu_p = - e \, \delta\varphi - e \, \alpha_p T(x) + e \, C_2.
\label{eq:dfi_dmun_dmup}
\end{equation}

Here
\[
\alpha ={\alpha_n n_0(T^*) + \alpha_p p_0(T^*) \over n_0(T^*) + p_0(T^*)}.
\]
Constants $A_{1,2}$ and $C_{1,2}$ [see (\ref{eq:dfi_dmun_dmup})]
are connected by correlations
\[
{n_0(T^*) C_1 - p_0(T^*) C_2 \over n_0(T^*) + p_0(T^*)} =  A_1 + A_2 -
							  \alpha T^*,
\]
\[
\xi_p C_2 - \xi_n C_1 = ( \alpha_p \xi_p + \alpha_n \xi_n) T_2 +
			  (\xi_n + \xi_p)\varphi_c
\]
[for definitions of $\xi_n$ and $\xi_p$, see (\ref{eq:lim})].

For determining all the constants presented in (\ref{eq:dfi_dmun_dmup}),
it is necessary to give conditions $\left. \delta\varphi\right|_{x=\pm
a}$ and $\delta\mu_n$ (or $\delta\mu_p$) at $x=0$, for
example.\cite{by_the_way}

Let [see (\ref{eq:surface_con_for_curr})]
\[
\left. \delta\varphi\right|_{x=\pm a} = \varphi_{c,b},\qquad
\delta\mu_n(0) = \delta\mu_p(0) = 0.
\]

Then expressions (\ref{eq:dfi_dmun_dmup}) turn into
\[
\delta\varphi = {\Delta T \over 2} \left[ { \alpha {x \over a} - \left(
 \alpha + {\alpha_n \xi_n + \alpha_p \xi_p \over \xi_n + \xi_p} \right)
 {\sinh {\lambda x} \over \sinh \lambda a} }\right],
\]
\[
\delta\mu_n = +e \, \delta \varphi - e \,\alpha_n {\Delta T }
 {x \over 2a},
\]
\begin{equation}
\delta\mu_p = -e \, \delta \varphi + e \,\alpha_p {\Delta T }
 {x \over 2a}.
\label{eq:new_dfi_dmun}
\end{equation}
As follows from formulas (\ref{eq:dfi_dmun_dmup}) and (\ref{eq:new_dfi_dmun}),
$\delta\mu_n \ne -\delta\mu_p$, e.g., two Fermi quasilevels really appear.

If we use the condition of quasineutrality $\lambda a \gg 1$, then
\[
\delta\varphi = \alpha{\Delta T}  {x \over 2a},
\]
\[
\delta\mu_n = e \,{p_0(T^*) \over n_0(T^*) + p_0(T^*)} (\alpha_p - \alpha_n)
	       {\Delta T } {x \over 2a},
\]
\[
\delta\mu_p = e\,{n_0(T^*) \over n_0(T^*) + p_0(T^*)} (\alpha_p - \alpha_n)
	       {\Delta T } {x \over 2a},
\]
\begin{equation}
\delta n = \delta p = e\,{n_0(T^*)p_0(T^*) \over n_0(T^*) + p_0(T^*)}
	   (\alpha_p - \alpha_n) {\Delta T \over T^*} {x \over 2a}.
\label{eq:quasi_new_dfi_dmun}
\end{equation}

Thus there are two Fermi quasilevels even in the quasineutrality
approximation and the condition of quasineutrality reduces to the
equality $\delta n = \delta p$ [compare with (\ref{eq:quasi_dfi1_dn1})].

When the semiconductor is unipolar [for example, $n_0(T^*) \gg p_0(T^*)$]
$\delta n = \delta p \ll p_0(T^*) \ll n_0(T^*)$ and $\delta\mu_n=0$,
e.g., the redistribution of concentration and Fermi level change do not
take place. This is in accordance with the results of the Introduction.

To conclude, let us note that expressions (\ref{eq:quasi_new_dfi_dmun})
could be derived from the equations ${\bf j}_n={\bf 0}$ and ${\bf
j}_p={\bf 0}$ only without using the Poisson equation, assuming at once
that the relation $\delta n=\delta p$ takes p lace when $\lambda a \gg 1$.

\section{Thermo-emf of bipolar semiconductor}
\label{sec:temfbps}
As was noted above, the thermo-emf is described by the expression
(\ref{eq:temf_of_any_nature}) in a bipolar
medium, and it is important to emphasize that this expression does not
contain the electric potential $\varphi(x)$. The latter is quite natural
if we wish to use the correct determination of the thermoemf, which is
formed by forces of nonelectric origin,
but not to use the artificial scheme which was presented in the first section.

Let us assume that a semiconductor sample is placed in the interval $-a
\le x \le +a$. It is connected with an instrument by metal wires with
chemical potential $\mu_M$ which does not depend on temperature. Let the
thermoelectric power $\alpha_M$ be equal
to zero ($\alpha_M \ll \alpha_n,\alpha_p$). We assume that
$\mu_n(T^*)=\mu_M$ for simplicity.

As was noted, the first integral in expression
(\ref{eq:temf_of_any_nature}) tends to zero
when electrons and holes have a single level of chemical potential
($\delta\mu_n=-\delta\mu_p$). So we have: \[
V = \oint \left( {\sigma_n \over e \, \sigma} {d \over dx} \delta \mu_n -
{\sigma_p \over e \, \sigma} {d \over dx} \delta \mu_p \right) dl -
\oint\left( {\sigma_n \alpha_n \over \sigma} + {\sigma_p
\alpha_p \over \sigma} \right) {dT \over dx}dl.
\]

Taking into account that $\delta\mu_n$ and $\delta\mu_p\propto\Delta T$ [and
so $\sigma_n$ and $\sigma_p$ depend on $n_0(T^*)$ and $p_0(T^*)$], we
get
\begin{eqnarray}
V &=& {1 \over e \sigma(T^*)} \left[ \sigma_n(T^*) \int_{-a}^{+a}
     d(\delta\mu_n) - \sigma_p(T^*) \int_{-a}^{+a} d(\delta \mu_p)
\right]   \nonumber \\
  & & \quad + {\xi_n \over e \, \xi} \left[ \delta\mu_n(-a) -
\delta\mu_n(+a)\right]
     - {\xi_p \over e \, \xi} \left[ \delta\mu_p(-a) -
\delta\mu_p(+a)\right] \nonumber \\
  & & \qquad - {1 \over \sigma(T^*)}
	 \left[ \sigma_n (T^*) \alpha_n
	       + \sigma_p(T^*) \alpha_p \right] (T_2 -T_1);\nonumber \\
\xi &=& \xi_n + \xi_p.
\label{eq:temf_bpsc}
\end{eqnarray}

The second and third terms in the expression (\ref{eq:temf_bpsc})
correspond to the contributions of the jumps of $\mu_n$ and $\mu_p$ on the
surfaces $x=\pm a$ to the first integral (\ref{eq:temf_of_any_nature}).
The analogous contribution of the second integral is absent, since in our
problem temperature is continuous at $x=\pm a$
[compare with (\ref{eq:surface_con_for_curr})]. Two additional terms
$[(\xi_n \alpha_n^s / \xi) + (\xi_p\alpha_p^s / \xi)]\Delta T_\pm$
appear in expressions (\ref{eq:surface_con_for_curr}) and
(\ref{eq:temf_bpsc}) if the thermoconductivity of
planes $x=\pm a$ is a finite quantity, and thus the temperature has
discontinuities $\Delta T_\pm$.\cite{BBG} The index ``$s$'' serves to show
the relation to the planes $x=\pm a$.

We find from expressions (\ref{eq:temf_bpsc}) and (\ref{eq:dfi_dmun_dmup}):
\begin{eqnarray}
V= {\xi_n \alpha_n + \xi_p \alpha_p \over \xi} \Delta T.
\label{eq:new_temf_bpsc}
\end{eqnarray}
For example, the Fermi quasilevel contribution to the thermo-emf
compensates completely
the conventional thermo-emf expression (\ref{eq:bpt_V}). The nonzero
thermo-emf is caused only by Fermi quasilevel breaks in the contacts of the
semiconductor sample with connecting wires. Let us emphasize once again
that the last assertion is true with any form of boundary conditions on
planes $x=\pm a$.

\section*{Conclusion}
Thus the thermo-emf $V$ is determined by thermoelectric powers
$\alpha_{n,p}$ and surface characteristics $\xi_{n,p}$. Comparing
formula (\ref{eq:new_temf_bpsc}) with formula (\ref{eq:bpt_V})
we see that taking into account Fermi quasilevels in the thermo-emf
changes its value substantially. So if $\alpha_n \sigma_n(T^*)
+ \alpha_p \sigma_p(T^*) =0$
then expression (\ref{eq:bpt_V}) becomes zero but the value $V$ obtained
from (\ref{eq:new_temf_bpsc}) does not equal zero.
In the general case expressions (\ref{eq:bpt_V}) and (\ref{eq:new_temf_bpsc})
may have different signs.

Returning to the unipolar case ($\xi_n \gg \xi_p$) we come to the expression
(\ref{eq:simp_unipol_V}) as was noted above. Let us notice only
that the condition $\xi_p \gg \xi_n$ does not follow from the condition
$p_0(T^*) \gg n_0(T^*)$ in a hole semiconductor.
The problem is that the ratio of the electron mobility to the hole mobility
can become very large in some semiconductors [for example, it is more than 80
in InSb (Ref.~\onlinecite{Ioffe60})], and the transition from expression
(\ref{eq:new_temf_bpsc}) to (\ref{eq:simp_unipol_V}) does not always take
place.

It is clear that use of expression (\ref{eq:new_temf_bpsc}) can
essentially change the calculated value of the efficiency of a thermoconverter.
We would like to finish with one comment. It is impossible to say anything
about the order of magnitude of $V$ in typical experimental situations, because
nobody knows the value of $\xi$.

\acknowledgments
This work was partially supported by Consejo Nacional de Ciencia y
Tecnologia (CONACyT-M\'{e}xico). The authors are grateful to M.~Shapiro
for stimulating discussions.

\appendix{}
\section{}
The chemical potential $\mu_n^0\biglb( T(x) \bigrb)$
[see (\ref{eq:temp_over_conditions})] corresponding to concentrations
$n_0(x)$
and $p_0(x)$ is heterogeneous in space. So diffusion
currents will
arise in the process of the establishment of ``thermodynamic equilibrium'',
which
leads to redistribution of concentrations, the appearance of bulk charge
layers, and an internal thermoelectric field characterized by the electric
potential $\delta\varphi_1(x)$. Here the situation is analogous to
the process of establishing thermodynamic equilibrium in heterogeneously doped
semiconductors.\cite{Anselm78}
The following distributions of the electric potential $\delta\varphi_1(x)$,
concentrations, and chemical potentials:
\[
n_1(x)=n_0(x) + \delta n_1(x),\qquad p_1(x)=p_0(x) +\delta p_1(x),
\]
\[
\mu_n^{(1)}(x)=\mu_n^0(x) + \delta\mu_n^{(1)}(x),\qquad
	\mu_p^{(1)}(x)=\mu_p^0(x) + \delta\mu_p^{(1)}(x),
\]
correspond to ``equilibrium'' when the electrochemical potential is constant:
\begin{equation}
\mu_n^{(1)}(x) - e\,\delta\varphi_1(x)
	=-\varepsilon_g - \mu_p^{(1)}(x) - e \, \delta\varphi_1(x)
	= {\rm const}.
\label{eq:chem_pot_cond}
\end{equation}

The functions $\delta n_1(x)$ and $\delta p_1(x)$ are connected with
$\delta\mu_n^{(1)}$ and $\delta\mu_p^{(1)}$ [see
(\ref{eq:temp_consentrations})~and~(\ref{eq:temp_con_of_holes})]
naturally by the formulas
\begin{equation}
\delta n_1=n_0(T^*){\delta\mu_n^{(1)} \over T^*},\qquad
\delta p_1=p_0(T^*){\delta\mu_p^{(1)} \over T^*},\qquad
\delta\mu_p^{(1)}=-\delta\mu_n^{(1)}.
\label{eq:3d}
\end{equation}

Thus there are two unknown independent functions. It is necessary to use
the Poisson equation to determine them, besides Eq. (\ref{eq:chem_pot_cond}).
\begin{equation}
{d^{\,2}(\delta\varphi_1) \over dx^2}=4\pi e(\delta n_1 - \delta p_1).
\label{eq:Poisson_eq}
\end{equation}

It is easy to formulate boundary conditions for
Eqs.~(\ref{eq:chem_pot_cond}) and (\ref{eq:Poisson_eq}) if we introduce
electron and hole electric conductances per unit area of the contacts of
the semiconductor sample with connecting wires
($\xi_n$ and $\xi_p$, respectively). If the thickness of the junction
$\delta$
is negligible compared to the thickness $2a$ of the bulk semiconductor and
$\sigma_n^s$ ($\sigma_p^s$) is the junction conductivity of electrons
(holes), we have ($\sigma_n^s$ and $\sigma_p^s$ are supposed to depend on
$\delta$)\cite{BBG}
\begin{equation}
\xi_n=\lim\limits_{\delta\to0}{\sigma_n^s \over \delta}
\text{ and }
\xi_p=\lim\limits_{\delta\to0}{\sigma_p^s \over \delta} .
\label{eq:lim}
\end{equation}
We choose the origin of $\varphi(x)$ at the point $x=0$ [$\left.
\varphi(x)\right|_{x=0}=0$].

Then the condition of continuity of electric current in contacts in the
broken circuit [considering that $\alpha_n=\alpha_p=0$ in the connecting
wires and $\mu_M$ does not depend on temperature (metal)] is reduced
to\cite{BBG}
\begin{equation}
\left.
{  {1 \over e} \xi_p\delta\tilde\mu_p - {1 \over e} \xi_n\delta\tilde\mu_n }
\right|_{x=\pm a} =(\xi_n + \xi_p)\varphi_{c,b}.
\label{eq:surface_con_for_curr}
\end{equation}

Condition~(\ref{eq:surface_con_for_curr}) in the ``thermodynamic
equilibrium state'' turns into \[
\left.
{ \delta\mu_n^0 + \delta\mu_n^{(1)} - e\,\delta\varphi_1}
\right|_{x=\pm a} =0,
\]
since $\varphi_c=\varphi_b=0$ and $\delta\mu_p=-\delta\mu_n$.

As a result, for $\delta\varphi_1$, $\delta\mu_n^{(1)}$
($\delta\mu_p^{(1)} = - \delta\mu_n^{(1)}$)
and $\delta n_1$  [$n_0(T^*)\, \delta p_1 =- p_0(T^*)\, \delta n_1$]
we find
\begin{eqnarray}
\delta\varphi_1 = {1 \over e}\delta\mu_n^{(0)}(x) + 2\,C \; \sinh \lambda x,
\nonumber\\
\delta\mu_n^{(1)} = 2 \, C \, e \; \sinh \lambda x,\\
\label{eq:3d2}
\delta n_1 = 2 \, C \, e {n_0(T^*) \over T^*} \; \sinh \lambda x.\nonumber
\end{eqnarray}
Here
\[
\lambda^{-1} =\left\{ {T^* \over 4\pi e^2 \left[ n_0(T^*) +
		p_0(T^*)\right]} \right\}^{1/2}
\]
is the Debye radius.

It is necessary to formulate boundary conditions either
on $\delta\varphi_1$ or on $\delta\mu_n^{(1)}$ for determining the constant
$C$. It is natural to assume that
$\left.{ \delta\varphi_1} \right|_{x=\pm a}=0$. Then
\[
    C={\partial \mu_n^0(T^*) \over \partial T^*}{\Delta T \over 2 \, e}
	   {1 \over 2 \; \sinh \lambda a} .
\]

If the condition of quasineutrality holds, e.g., $\lambda a \gg 1$,
as usual, takes place, then
\begin{eqnarray}
\delta\varphi_1 &=& -{d\mu_n^0(T^*) \over dT^*} {\Delta T \over e}
			{x \over 2a},\nonumber \\
\delta n_1 &=& \delta p_1 = \delta\mu_n^{(1)} = \delta\mu_p^{(1)} =0.
\label{eq:quasi_dfi1_dn1}
\end{eqnarray}

At first sight it seems that Eq.~(\ref{eq:Poisson_eq}) implies the
condition $\delta n_1=\delta p_1$ only.   But we see
from~(\ref{eq:quasi_dfi1_dn1}) that the condition of quasineutrality
reduces to the stronger requirement $\delta n_1=\delta p_1=0$ in a
bipolar medium as in a unipolar semiconductor during the process of
establishing ``equilibrium.''

It is clear that functions $n_1(x)$ and $p_1(x)$ have to be named the
``equilibrium'' concentrations of the carriers because \[
{d \over dx}\left[ \mu_n^{(1)}(x) - e\,\delta\varphi_1(x) \right] =0,\qquad
{d \over dx}\left[ \mu_p^{(1)}(x) + e\,\delta\varphi_1(x) \right] =0.
\]
These functions must participate exactly as ``equilibrium''
concentrations in the expressions for the bulk and surface recombinations.

\begin{figure}
\caption{The electric circuit for measuring of thermo-emf.}
\label{fig:single}
\end{figure}
\newpage
\begin{figure}
\begin{center}\begin{picture}(300,125)(-150,0)
\put(   0, 25){\circle{20}}
\put(- 13, 12){\vector(1,1){26}}
\put(-100, 25){\line( 1,0){90}}
\put( 100, 25){\line(-1,0){90}}
\put(-100, 35){\oval(20,20)[lb]}
\put( 100, 35){\oval(20,20)[rb]}
\put(-110, 35){\line(0,1){55}}
\put( 110, 35){\line(0,1){55}}
\put(-100, 90){\oval(20,20)[lt]}
\put( 100, 90){\oval(20,20)[rt]}
\put(-100,100){\line( 1,0){50}}
\put( 100,100){\line(-1,0){50}}
\put(- 50, 85){\framebox(100,30){semiconductor}}
\put(- 70, 85){\vector(1,0){140}}
\put(   0, 82){\line(0,1){6}}
\put(- 55, 75){$-a$} \put( 45,75){$a$}   \put(  0,75){$0$}\put(75,75){$x$}
\put(- 55,120){$T_1$}\put( 45,120){$T_2$}
\put(- 55, 30){$T^*$}\put( 45, 30){$T^*$}
\put(- 55, 10){$b$}  \put( 45, 10){$c$}
\put(- 50, 25){\circle*{3}}  \put( 50, 25){\circle*{3}}
\put(- 95,-10){FIG.\ 1. Yu.\ G. Gurevich, Phys.\ Rew.\ B}
\end{picture}\end{center}
\end{figure}

\begin{references}
\bibitem[*]{byline}Permanent address: Institute for Radiophysics
	and Electronics, Academy of Science of Ukraine, Kharkov 310085,
	Ukraine.
\bibitem{Anselm78} A.~I.~Anselm,
        {\it Introduction to semiconductor theory}
        (Mir, Moscow, Prentice-Hall, Englewood Cliffs, NJ, 1981).

\bibitem{Thompson1882} W. Thomson, {\it Mathematical and physical papers}
	(Cambridge University Press, Cambridge, England, 1882), Vol. 1, p. 266.

\bibitem{Kaydanov81} K.~Seeger, {\it Semiconductor physics} (Springer-Verlag,
	New York, 1973);
	V.~I.~Kaydanov and A.~V.~Nuramski,
	{\it Electric conductivity,
	thermoelectric phenomena and thermal conductivity of semiconductors}
        (LPI, Leningrad, 1981).

\bibitem{GY} Yu.~G.~Gurevich and V.~B.~Yurchenko,
	Fiz.\ Tekh.\ Poluprovodn.\ {\bf 25}, 2109 (1991)
	[Sov.\ Phys.\ Semicond. {\bf 25}, 1268  (1991)].

\bibitem{Ioffe60} A.~F.~Ioffe,
	{\it Semiconductor thermo-elements}
	(Izd.\ Ak.\ Nauk USSR, Moscow, 1960).  

\bibitem{BBG} F.~G.~Bass, V.~S.~Bochkov, and Yu.~G.~Gurevich,
	{\it Electrons and phonons in limit semiconductors}
        (Nauka, Moscow, 1984).

\bibitem{GM} Yu.~G.~Gurevich and O.~L.~Mashkevich,
	Fiz.\ Tekh.\ Poluprovodn.\ {\bf 24}, 1327 (1990)
	[Sov.\ Phys.\ Semicond.\ {\bf 24},  835 (1990)].

\bibitem{GL} Yu.~G.~Gurevich  and G.~N.~Logvinov,
	Phys.\ Rev.\ B {\bf 46}, 15 516  (1992).
\bibitem{by_the_way}It is important to emphasize
(see Sec.~\ref{sec:temfbps}) that constants $C_{1,2}$ and $A_{1,2}$ are
displaced out from the answer under calculation of the thermo-emf, e.g.,
the value of the thermo-emf does not depend on the concrete form of the
boundary conditions for $\delta\varphi (x)$ and
$\delta\mu_{n,p}(x)$.

\end{references}
\end{document}